\documentclass{article}

\usepackage[final]{neurips_2025_ml4ps}

\usepackage[utf8]{inputenc} 
\usepackage[T1]{fontenc}    
\usepackage{url}            
\usepackage{booktabs}       
\usepackage{amsfonts}       
\usepackage{nicefrac}       
\usepackage{microtype}      
\usepackage{xcolor}         
\usepackage{amsmath}
\usepackage{mathtools}
\usepackage{graphicx}
\usepackage{tikz}
\usepackage{varioref}
\usepackage[hidelinks]{hyperref}
\usepackage{cleveref}
\crefname{figure}{Fig.}{Figs.}
\crefname{table}{Tab.}{Tabs.}
\crefname{section}{Sec.}{Secs.}
\crefname{appendix}{Appendix}{Appendices}
\usepackage{xspace}
\usepackage{chngcntr}
\usepackage{fontawesome5}

\newcommand{\jax}{\texttt{jax}\xspace}
\newcommand{\jaxpm}{\texttt{JaxPM}\xspace}

\newcommand{\hyper}{\texttt{HYPER}\xspace}
\newcommand{\diffhydro}{\texttt{diffHydro}\xspace}
\newcommand{\diffrax}{\texttt{diffrax}\xspace}
\newcommand{\flax}{\texttt{flax}\xspace}
\newcommand{\optax}{\texttt{optax}\xspace}

\newcommand{\camels}{\textsc{Camels}\xspace}
\newcommand{\simba}{\textsc{Simba}\xspace}
\newcommand{\astrid}{\textsc{Astrid}\xspace}
\newcommand{\illustris}{\textsc{IllustrisTNG}\xspace}

\newcommand{\lcdm}{\ensuremath{\Lambda}CDM\xspace}

\newcommand{\omatter}{\ensuremath{\Omega_\mathrm{m}}\xspace}
\newcommand{\ocdm}{\ensuremath{\Omega_\mathrm{cdm}}\xspace}
\newcommand{\obary}{\ensuremath{\Omega_\mathrm{b}}\xspace}
\newcommand{\sigeight}{\ensuremath{\sigma_8}\xspace}

\newcommand{\ns}{\ensuremath{n_s}\xspace}

\newcommand{\dif}{\mathrm{d}}

\title{Hybrid Physical-Neural Simulator for \\Fast Cosmological Hydrodynamics}

\author{
  Arne Thomsen \\
  ETH Zürich \\
  Zürich, Switzerland \\
  athomsen@phys.ethz.ch
  \And
  Tilman Tröster \\
  ETH Zürich \\
  Zürich, Switzerland
  \And
  François Lanusse \\
  Université Paris-Saclay \\
  CEA, CNRS, AIM \\
  Gif-sur-Yvette, France
}

\begin{document}

\maketitle

\begin{abstract}
    Cosmological field-level inference requires differentiable forward models that solve the challenging dynamics of gas and dark matter under hydrodynamics and gravity. 
    We propose a hybrid approach where gravitational forces are computed using a differentiable particle-mesh solver, while the hydrodynamics are parametrized by a neural network that maps local quantities to an effective pressure field.
    We demonstrate that our method improves upon alternative approaches, such as an Enthalpy Gradient Descent baseline, both at the field and summary-statistic level.
    The approach is furthermore highly data efficient, with a single reference simulation of cosmological structure formation being sufficient to constrain the neural pressure model.
    This opens the door for future applications where the model is fit directly to observational data, rather than a training set of simulations.~\href{https://github.com/Arne-Thomsen/JaxHPM}{\faGithub}
    \end{abstract}

\section{Introduction}\label{sec:introduction}
Analyses of the large-scale structure of the Universe aim to constrain fundamental cosmological parameters. 
While structure formation at large scales is dominated by dark matter, state-of-the-art observations now probe smaller scales where the contribution of ordinary (baryonic) matter in the form of gas becomes non-negligible, necessitating hydrodynamical modeling.

Explicit field-level inference leverages the full spatial information in cosmic density and velocity fields to jointly constrain cosmological parameters and initial conditions of the matter distribution~\citep{jascheBayesianPhysicalReconstruction2013}. 
Because the parameter space of initial conditions is extremely high-dimensional, their inference requires gradient-based sampling methods like Hamiltonian Monte Carlo~\citep{betancourtConceptualIntroductionHamiltonian2018} and hence differentiable forward models.

We present a step towards such a differentiable simulation jointly evolving gas and dark matter.
Recognizing that the majority of cosmic matter is dark matter, whose purely gravitational dynamics can be solved efficiently with differentiable particle mesh (PM) methods~\citep{hockneyComputerSimulationUsing1988}, we propose a hybrid approach that maintains physics-driven gravitational dynamics while approximating gas physics through a physically-constrained neural network parametrization.
We demonstrate this solver-in-the-loop scheme~\citep{umSolverintheLoopLearningDifferentiable2021} by fitting the trainable network parameters to a single fully hydrodynamical reference simulation.

\section{Related work}\label{sec:related_work}

\paragraph{HYPER}
The hydro PM (HPM) code \hyper introduced by~\citet{heHydroParticleMeshCodeEfficient2022} shares conceptual similarities with our approach, modeling the evolution of both dark matter and gas particles within a PM framework where additional gas forces are derived from a scalar pressure field.
However, unlike in our data-driven parametrization, the functional form of the pressure field in HPM is derived analytically using the halo model~\citep{asgariHaloModelCosmology2023}, with distinct treatments for the lower-density intergalactic and higher-density intracluster medium.

\paragraph{diffHydro}
The \diffhydro code~\citep{horowitzDifferentiableCosmologicalHydrodynamics2025} implements a fully differentiable approach to cosmological hydrodynamical simulations. 
In contrast to the hybrid method presented in this work, \diffhydro is purely physics-driven, solving the hydrodynamical Euler equations in comoving coordinates using a finite volume scheme, at significant computational cost.

\paragraph{Enthalpy Gradient Descent (EGD)}
The EGD method developed in~\citet{daiGradientBasedMethod2018} is a fast, differentiable technique for post-processing dark matter-only simulations to include gas effects.
Under the assumption of an effective power-law equation of state $T(\delta) = T_0(1 + \delta)^{(\gamma -1)}$, the method constructs an enthalpy field from the matter density contrast $\delta$ and displaces a random subset of dark matter particles along the enthalpy gradient to approximate gas dynamics.
The free parameters $T_0$ and $\gamma$ can be fit to reference hydrodynamical simulations by optimizing the matter power spectrum.

For this work, we use EGD as our primary baseline comparison since neither \hyper nor \diffhydro have publicly available implementations.

\section{Method}\label{sec:method}
We extend the open-source dark matter-only PM code \jaxpm\footnote{\url{https://github.com/DifferentiableUniverseInitiative/JaxPM}} from~\citet{lanzieriHybridPhysicalNeuralODEs2022} by introducing a gas particle species that, in addition to gravity, experiences a learned pressure force.
As a PM code, forces derived from gradients of scalar fields can be efficiently computed through point-wise multiplications in Fourier space, with cloud-in-cell interpolation to transition between particle and mesh representations.
The \jax~\citep{jax2018github} implementation provides the automatic differentiation, just-in-time compilation, and GPU acceleration necessary to train the neural pressure model.

\subsection{Hybrid physical-neural equations of motion}\label{sec:equations_of_motion}
The simulator evolves equal numbers of cold dark matter and baryonic gas particles with masses proportional to the cosmological density fractions \ocdm and \obary, respectively.
Similar to~\citet{heHydroParticleMeshCodeEfficient2022}, we integrate the system of ordinary differential equations (ODEs) in comoving coordinates:
\begin{equation}
\left\{
    \begin{aligned}
        \frac{\dif \mathbf{x}_\mathrm{dm}}{\dif a} &= \frac{1}{a^3 E(a)} \mathbf{v}_\mathrm{dm} \\
        \frac{\dif \mathbf{v}_\mathrm{dm}}{\dif a} &= - \frac{1}{a^2 E(a)} \nabla \Phi_\mathrm{tot},
    \end{aligned}
\right.
\qquad
\left\{
\begin{aligned}
        \frac{\dif \mathbf{x}_\mathrm{gas}}{\dif a} &= \frac{1}{a^3 E(a)} \mathbf{v}_\mathrm{gas} \\
        \frac{\dif \mathbf{v}_\mathrm{gas}}{\dif a} &= - \frac{1}{a^2 E(a)} \left(\nabla \Phi_\mathrm{tot} + \frac{\nabla P}{\rho_\text{gas}}\right),
    \end{aligned}
\right.
\label{eq:ode}
\end{equation}
where $\mathbf{x}$ and $\mathbf{v}$ denote particle positions and velocities, $a$ is the cosmological scale factor serving as the time variable, and $E(a) = H(a)/H_0$ is the dimensionless Hubble parameter.
The scale factor-dependent prefactors account for the background expansion of the universe~\citep{quinnTimeSteppingNbody1997,schallerSwiftModernHighly2024}.
The force terms proportional to $- \nabla \Phi_\mathrm{tot}$ and $-\nabla P/\rho_\mathrm{gas}$ are detailed below.

The dynamics of both particle species are coupled through the scalar gravitational potential $\Phi_\mathrm{tot}$, which is related to the total matter density contrast $\delta_\mathrm{tot}$ via the Poisson equation~\citep{anguloLargescaleDarkMatter2022}
\begin{equation}
    \nabla^2 \Phi_\mathrm{tot}(\mathbf{x}) = \frac{3}{2} H_0^2 \omatter \delta_\mathrm{tot}(\mathbf{x}),
    \label{eq:potential}
\end{equation} 
where $\delta_\mathrm{tot} = \delta_\mathrm{dm} + \delta_\mathrm{gas}$ and $\delta = \rho / \bar{\rho} - 1$ with $\rho$ the local density and $\bar{\rho}$ its spatial mean.
Here, $\omatter = \ocdm + \obary$ is the total matter density fraction.
The density fields are computed on a mesh, enabling efficient solution of \cref{eq:potential} in Fourier space.
Adopting the approach of~\citet{lanzieriHybridPhysicalNeuralODEs2022}, we optionally include a residual neural correction $f_\theta$ to the Fourier-transformed potential: $\tilde{\Phi}_\mathrm{tot}^\star = (1 + f_\theta(a, |\mathbf{k}|)) \, \tilde{\Phi}_\mathrm{tot}$.
This hybrid approach combines the physics-driven potential with a data-driven correction to improve small-scale accuracy.

\subsection{Effective neural pressure}\label{sec:neural_pressure}
The effective pressure force acting on gas particles takes the physically motivated Euler form $- \nabla P / \rho_\mathrm{gas}$ as defined in \cref{eq:ode}.
Under an ideal gas law assumption, we express the learned pressure field at particle position $\mathbf{x}$ as
\begin{equation}
    P_\varphi(a, \mathbf{x}) \propto \rho_\mathrm{gas}(\mathbf{x}) \, U_\varphi(a, \mathbf{h}(\mathbf{x)}),
    \label{eq:pressure}
\end{equation}
where $U_\varphi$ is a neural network with trainable parameters $\varphi$ that maps the scale factor $a$ and feature vector of local quantities $\mathbf{h}(\mathbf{x})$ to the non-negative internal energy. 
This decomposition reduces the dynamic range compared to predicting $P$ directly, significantly improving training stability.
We implement $U_\varphi$ as a fully convolutional neural network (CNN) operating on the mesh interpolation, though the framework supports alternative architectures such as multilayer perceptrons (MLPs) acting directly on particles.
Both architectures operate locally, decoupling the box sizes of reference and hybrid simulations and hence enabling future application to larger cosmological volumes.

For computational efficiency within the PM framework, we construct input features $\mathbf{h}(\mathbf{x})$ that can be rapidly evaluated on the mesh.
The resulting feature vector comprises
\begin{equation}
    \mathbf{h}(\mathbf{x}) = \left(\rho_\mathrm{gas}(\mathbf{x}), \, f_\mathrm{scalar}(\mathbf{x}), \, \nabla_{\! \mathbf{v}}(\mathbf{x}), \, \sigma_\mathbf{v}^2(\mathbf{x})\right),
    \label{eq:features}
\end{equation}
where $\rho_\mathrm{gas}$ is the gas density, $\nabla_{\! \mathbf{v}} = \nabla \cdot \mathbf{v}$ is the velocity divergence, and $\sigma_\mathbf{v}^2 = |\mathbf{v} - \langle\mathbf{v}\rangle|^2$ is the velocity dispersion with $\langle\mathbf{v}\rangle$ denoting the local mean velocity.
As in~\citet{heHydroParticleMeshCodeEfficient2022}, we define 
\begin{equation}
    f_\mathrm{scalar}(\mathbf{x}) \propto \rho_\mathrm{gas}(\mathbf{x}) \ast \frac{1}{|\mathbf{x}|^2} \Rightarrow \tilde{f}_{\mathrm{scalar}}(\mathbf{k}) \propto \frac{2\pi^2 \tilde{\rho}_\mathrm{gas}(\mathbf{k})}{|\mathbf{k}|},
    \label{eq:fscalar}
\end{equation}
which we efficiently compute in Fourier space.
The input fields are visualized in~\cref{fig:hpm_inputs}.
We empirically find that extending $\mathbf{h}(\mathbf{x})$ by the rotationally invariant eigenvalues of the tidal field tensor (or Hessian of the gravitational potential) does not significantly improve performance and therefore do not include them.

\begin{figure}[t]
    \centering
    \includegraphics[width=\linewidth]{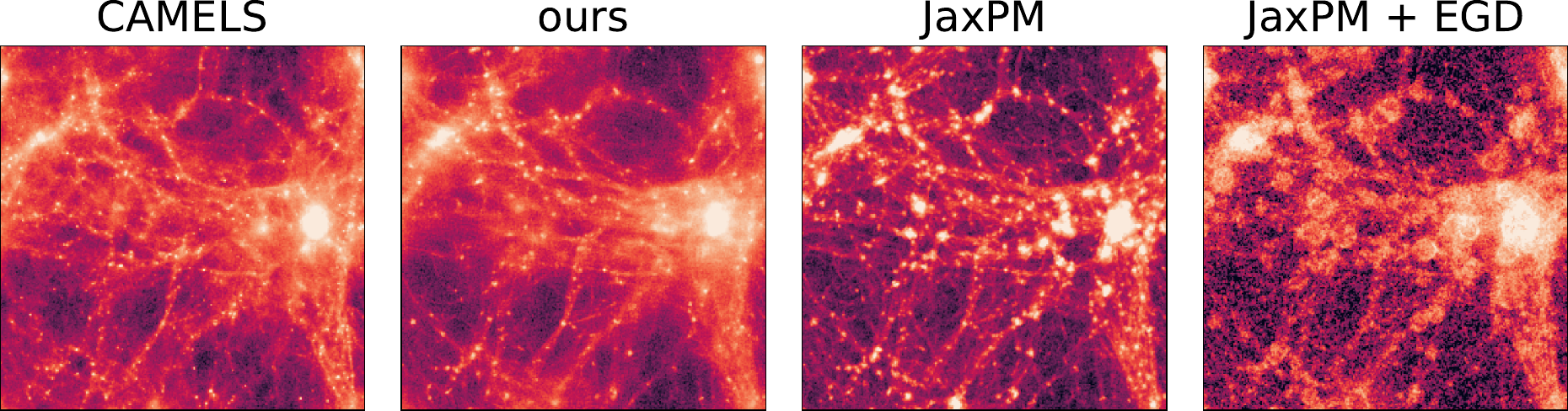}
    \caption{
        Projected gas density $\rho_\mathrm{gas}$ (logarithmic scale) comparing four methods: 
        reference hydrodynamical \camels simulation, our hybrid simulator with learned gas pressure force, gravity-only \jaxpm, and its EGD post-processing.
        Our pressure model suppresses small-scale structure formation, better matching the hydrodynamical reference than purely gravitational evolution while retaining more detail than EGD.
    }
    \label{fig:final_fields}
\end{figure}

We integrate \cref{eq:ode} using \diffrax\footnote{\url{https://docs.kidger.site/diffrax/}}~\citep{kidgerNeuralDifferentialEquations2022}, which allows automatic differentiation with respect to model parameters $\varphi$ using the recursive checkpoint adjoint~\citep{wang2009minimal,stumm2010new}.
Following~\citet{lanzieriHybridPhysicalNeuralODEs2022}, we exploit this differentiability to optimize our hybrid solver-in-the-loop~\citep{umSolverintheLoopLearningDifferentiable2021} model directly against a reference simulation (see \cref{sec:camels}) with particle positions $\mathbf{x}^\mathrm{ref}$ and velocities $\mathbf{v}^\mathrm{ref}$.
We minimize the loss function
\begin{equation}
    \mathcal{L} = \sum_s \left[ H_\delta\left(\mathbf{r}_s\right) + \lambda H_{\delta'}\left(\mathbf{v}_s - \mathbf{v}_s^\mathrm{ref}\right) + \mu \left\| \frac{P_s(|\mathbf{k}|)}{P_s^\mathrm{ref}(|\mathbf{k}|)} - 1 \right\|_2^2 \right],
    \label{eq:loss}
\end{equation}
where $H_\delta(r) = \min(r^2/2, \delta|r| - \delta^2/2)$ is the robust Huber loss that reduces sensitivity to outlier particles, $\mathbf{r}_s = \left(\left(\mathbf{x}_s - \mathbf{x}_s^\mathrm{ref} + L/2\right) \bmod L\right) - L/2$ denotes the particle displacement under periodic boundary conditions in a simulation box of side length $L$, $P(|\mathbf{k}|)$ is the power spectrum of $\rho_\mathrm{gas}$, and $\{\lambda,\mu,\delta, \delta'\}$ are hyperparameters.
The index $s$ runs over simulation snapshots (or a sparse subset thereof) taken at different scale factors $a_s$ during cosmic evolution. 
Example reference snapshots are shown in the top row of \cref{fig:appendix_evolution}.

\section{Results}\label{sec:results}
We demonstrate the data efficiency of our hybrid approach by training the neural pressure model on a single fully hydrodynamical reference simulation from the \simba\footnote{\url{http://simba.roe.ac.uk/}} subset~\citep{daveSimbaCosmologicalSimulations2019} of the  \camels\footnote{\url{https://camels.readthedocs.io/en/latest/index.html}} suite~\citep{villaescusa-navarroCAMELSProjectCosmology2021}, detailed in \cref{sec:camels}.
We select \simba because its strong gas feedback provides a more challenging test than \astrid~\citep{birdASTRIDSimulationGalaxy2022,niASTRIDSimulationEvolution2022}, whose weaker feedback produces negligible deviations from gravity-only dynamics at our resolution.

We downsample this reference to $128^3$ particles each for dark matter and gas, with an equally sized mesh for PM force computation, enabling training on an individual A100 GPU.
The system in \cref{eq:ode} is initialized using positions and velocities from the first \camels snapshot ($s = 0$) at scale factor $a_0 = 0.14$, corresponding to redshift $z_0 = 6$. 
Additional implementation details are provided in \cref{sec:network_implementation}.  

To further emphasize data efficiency, we evaluate the loss in \cref{eq:loss} at only 4 of the 34 available snapshots with scale factors $a_s \in \{0.30, 0.44, 0.65, 1.0\}$.
Training remains feasible despite this sparse supervision due to the physically constrained parametrization of the gas pressure force in \cref{eq:ode,eq:pressure}, the large number of simulated particles per snapshot, and the mesh size (128 cells per dimension) substantially exceeding the network's receptive field (17 cells).
Furthermore, the physics-driven gravitational dynamics in our hybrid approach provide a data-independent backbone, with gas physics acting as a learned correction.

All results are evaluated on held-out test simulations from \camels with different random initial conditions, ensuring model performance reflects generalization to cosmic variance rather than memorization of a specific realization.

\begin{figure}[h]
    \centering
    \includegraphics[width=\linewidth]{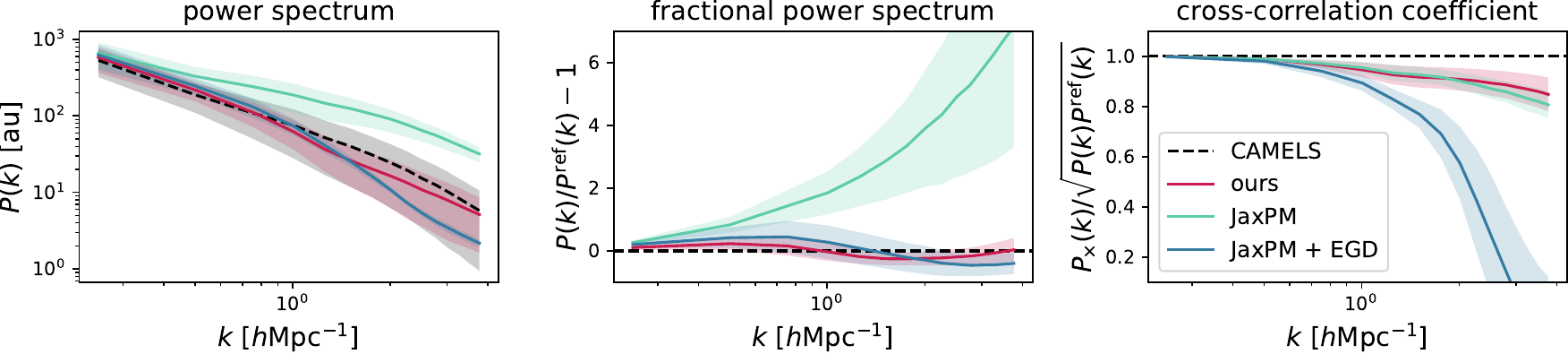}
    \caption{
        Two-point statistics of gas densities $\rho_\mathrm{gas}$ from \cref{fig:final_fields}.
        Solid lines and shaded bands indicate means and standard deviations over random initial conditions (cosmic variance), respectively.
        \textit{Left and middle:} Power spectra demonstrate that both our method and EGD suppress small-scale power relative to gravity-only evolution (\jaxpm), achieving better agreement with the \camels reference.
        \textit{Right:} Our hybrid simulator shows consistently higher cross-correlation with the reference than EGD.
        These quantitative results confirm the visual intuition from \cref{fig:final_fields}.
    }
    \label{fig:final_2pt}
\end{figure}

We present results at $a=1$ (present day), showing the density fields in \cref{fig:final_fields} and their corresponding two-point statistics measured across varying random initial conditions in \cref{fig:final_2pt}. 
We compare the reference \camels simulation against our hybrid method, exclusively gravitational \jaxpm evolution, and the EGD baseline (see \cref{sec:egd_implementation} for implementation details).
A key distinction exists between our method and this baseline: 
EGD post-processes gravity-only simulations at a single target scale factor, whereas our approach self-consistently evolves the system throughout cosmic history, even when trained on only a small subset of simulation snapshots. 
Results at different points in the evolution are shown in \cref{fig:appendix_evolution}.

We find that our hybrid extension significantly enhances gravity-only \jaxpm evolution at both the field and two-point levels.
Compared to EGD, our approach achieves similar improvement in the power spectrum (measuring Fourier mode amplitudes) while demonstrating superior field-level performance, as shown in \cref{fig:final_fields} and in the cross-correlation coefficient (capturing Fourier phase coherence).

Additionally, we show the dark matter distributions in \cref{fig:appendix_dm}, demonstrating negligible differences in baryonic back-reaction on dark matter across methods.

\section{Conclusion}\label{sec:conclusion}
We present a computationally efficient hybrid simulator that jointly evolves gas and dark matter particles by computing gravitational forces through fast PM methods while modeling gas pressure forces with an embedded physically constrained neural network.

This fully differentiable approach enables field-level inference, jointly constraining high-dimensional initial conditions with cosmological parameters. 
The method's data efficiency, requiring only a few snapshots of a single reference simulation for training, in principle opens the possibility of fitting directly to observational data rather than relying on extensive simulation training sets. 
For example, observations of gas, such as Sunyaev-Zeldovich effects, together with tomographic weak gravitational lensing could provide similar information on the gas and dark matter density as the sparse sampling of the simulations employed here. 
This could go towards addressing the recognized challenge of model misspecification between different hydrodynamical codes~\citep[e.g.][]{niCAMELSProjectExpanding2023,akhmetzhanovaDetectingModelMisspecification2025}.

Our neural parametrization of the pressure field uses only instantaneous local quantities derived from the particle positions and velocities, ignoring their history.
Complete thermodynamic treatment requires tracking an additional state variable like internal energy, entropy, or temperature.
This limitation could be overcome by augmenting \cref{eq:ode} with latent variables whose time derivatives are predicted by the embedded network, yielding a neural ODE~\citep{chenNeuralOrdinaryDifferential2019} that enables history-aware pressure predictions.

\bibliography{flatiron_hpm}

\begin{thebibliography}{28}
\providecommand{\natexlab}[1]{#1}
\providecommand{\url}[1]{\texttt{#1}}
\expandafter\ifx\csname urlstyle\endcsname\relax
  \providecommand{\doi}[1]{doi: #1}\else
  \providecommand{\doi}{doi: \begingroup \urlstyle{rm}\Url}\fi

\bibitem[Akhmetzhanova et~al.(2025)Akhmetzhanova, {Cuesta-Lazaro}, and {Mishra-Sharma}]{akhmetzhanovaDetectingModelMisspecification2025}
Aizhan Akhmetzhanova, Carolina {Cuesta-Lazaro}, and Siddharth {Mishra-Sharma}.
\newblock Detecting {{Model Misspecification}} in {{Cosmology}} with {{Scale-Dependent Normalizing Flows}}, August 2025.

\bibitem[Angulo and Hahn(2022)]{anguloLargescaleDarkMatter2022}
Raul~E. Angulo and Oliver Hahn.
\newblock Large-scale dark matter simulations.
\newblock \emph{Living Reviews in Computational Astrophysics}, 8\penalty0 (1):\penalty0 1, February 2022.
\newblock ISSN 2365-0524.
\newblock \doi{10.1007/s41115-021-00013-z}.

\bibitem[Asgari et~al.(2023)Asgari, Mead, and Heymans]{asgariHaloModelCosmology2023}
Marika Asgari, Alexander~J. Mead, and Catherine Heymans.
\newblock The halo model for cosmology: A pedagogical review.
\newblock \emph{The Open Journal of Astrophysics}, 6, November 2023.
\newblock \doi{10.21105/astro.2303.08752}.

\bibitem[Ba et~al.(2016)Ba, Kiros, and Hinton]{baLayerNormalization2016}
Jimmy~Lei Ba, Jamie~Ryan Kiros, and Geoffrey~E. Hinton.
\newblock Layer {{Normalization}}, July 2016.

\bibitem[Betancourt(2018)]{betancourtConceptualIntroductionHamiltonian2018}
Michael Betancourt.
\newblock A {{Conceptual Introduction}} to {{Hamiltonian Monte Carlo}}, July 2018.

\bibitem[Bird et~al.(2022)Bird, Ni, Di~Matteo, Croft, Feng, and Chen]{birdASTRIDSimulationGalaxy2022}
Simeon Bird, Yueying Ni, Tiziana Di~Matteo, Rupert Croft, Yu~Feng, and Nianyi Chen.
\newblock The {{ASTRID}} simulation: Galaxy formation and reionization.
\newblock \emph{Monthly Notices of the Royal Astronomical Society}, 512\penalty0 (3):\penalty0 3703--3716, May 2022.
\newblock ISSN 0035-8711.
\newblock \doi{10.1093/mnras/stac648}.

\bibitem[Bradbury et~al.(2018)Bradbury, Frostig, Hawkins, Johnson, Leary, Maclaurin, Necula, Paszke, VanderPlas, {Wanderman-Milne}, and Zhang]{jax2018github}
James Bradbury, Roy Frostig, Peter Hawkins, Matthew~James Johnson, Chris Leary, Dougal Maclaurin, George Necula, Adam Paszke, Jake VanderPlas, Skye {Wanderman-Milne}, and Qiao Zhang.
\newblock {{JAX}}: Composable transformations of {{Python}}+{{NumPy}} programs, 2018.

\bibitem[Chen et~al.(2019)Chen, Rubanova, Bettencourt, and Duvenaud]{chenNeuralOrdinaryDifferential2019}
Ricky T.~Q. Chen, Yulia Rubanova, Jesse Bettencourt, and David Duvenaud.
\newblock Neural {{Ordinary Differential Equations}}, December 2019.

\bibitem[Dai et~al.(2018)Dai, Feng, and Seljak]{daiGradientBasedMethod2018}
Biwei Dai, Yu~Feng, and Uro{\v s} Seljak.
\newblock A gradient based method for modeling baryons and matter in halos of fast simulations.
\newblock \emph{Journal of Cosmology and Astroparticle Physics}, 2018\penalty0 (11):\penalty0 009, November 2018.
\newblock ISSN 1475-7516.
\newblock \doi{10.1088/1475-7516/2018/11/009}.

\bibitem[Dav{\'e} et~al.(2019)Dav{\'e}, {Angl{\'e}s-Alc{\'a}zar}, Narayanan, Li, Rafieferantsoa, and Appleby]{daveSimbaCosmologicalSimulations2019}
Romeel Dav{\'e}, Daniel {Angl{\'e}s-Alc{\'a}zar}, Desika Narayanan, Qi~Li, Mika~H Rafieferantsoa, and Sarah Appleby.
\newblock Simba: {{Cosmological}} simulations with black hole growth and feedback.
\newblock \emph{Monthly Notices of the Royal Astronomical Society}, 486\penalty0 (2):\penalty0 2827--2849, June 2019.
\newblock ISSN 0035-8711.
\newblock \doi{10.1093/mnras/stz937}.

\bibitem[He et~al.(2016)He, Zhang, Ren, and Sun]{heDeepResidualLearning2016}
Kaiming He, Xiangyu Zhang, Shaoqing Ren, and Jian Sun.
\newblock Deep {{Residual Learning}} for {{Image Recognition}}.
\newblock In \emph{2016 {{IEEE Conference}} on {{Computer Vision}} and {{Pattern Recognition}} ({{CVPR}})}, pages 770--778, Las Vegas, NV, USA, June 2016. IEEE.
\newblock ISBN 978-1-4673-8851-1.
\newblock \doi{10.1109/CVPR.2016.90}.

\bibitem[He et~al.(2022)He, Trac, and Gnedin]{heHydroParticleMeshCodeEfficient2022}
Yizhou He, Hy~Trac, and Nickolay~Y. Gnedin.
\newblock A {{Hydro-Particle-Mesh Code}} for {{Efficient}} and {{Rapid Simulations}} of the {{Intracluster Medium}}.
\newblock \emph{The Astrophysical Journal}, 925\penalty0 (2):\penalty0 134, February 2022.
\newblock ISSN 0004-637X, 1538-4357.
\newblock \doi{10.3847/1538-4357/ac3bcb}.

\bibitem[Hockney and Eastwood(1988)]{hockneyComputerSimulationUsing1988}
R.~W. Hockney and J.~W. Eastwood.
\newblock \emph{Computer {{Simulation Using Particles}}}.
\newblock CRC Press, Boca Raton, 1988.
\newblock ISBN 978-0-367-80693-4.
\newblock \doi{10.1201/9780367806934}.

\bibitem[Horowitz and Lukic(2025)]{horowitzDifferentiableCosmologicalHydrodynamics2025}
Benjamin Horowitz and Zarija Lukic.
\newblock Differentiable {{Cosmological Hydrodynamics}} for {{Field-Level Inference}} and {{High Dimensional Parameter Constraints}}, February 2025.

\bibitem[Jasche and Wandelt(2013)]{jascheBayesianPhysicalReconstruction2013}
Jens Jasche and Benjamin~D. Wandelt.
\newblock Bayesian physical reconstruction of initial conditions from large-scale structure surveys.
\newblock \emph{Monthly Notices of the Royal Astronomical Society}, 432\penalty0 (2):\penalty0 894--913, June 2013.
\newblock ISSN 0035-8711.
\newblock \doi{10.1093/mnras/stt449}.

\bibitem[Kidger(2022)]{kidgerNeuralDifferentialEquations2022}
Patrick Kidger.
\newblock On {{Neural Differential Equations}}, February 2022.

\bibitem[Kingma and Ba(2017)]{kingmaAdamMethodStochastic2017}
Diederik~P. Kingma and Jimmy Ba.
\newblock Adam: {{A Method}} for {{Stochastic Optimization}}, January 2017.

\bibitem[Lanzieri et~al.(2022)Lanzieri, Lanusse, and Starck]{lanzieriHybridPhysicalNeuralODEs2022}
Denise Lanzieri, Fran{\c c}ois Lanusse, and Jean-Luc Starck.
\newblock Hybrid {{Physical-Neural ODEs}} for {{Fast N-body Simulations}}, July 2022.

\bibitem[Nelson et~al.(2019)Nelson, Springel, Pillepich, {Rodriguez-Gomez}, Torrey, Genel, Vogelsberger, Pakmor, Marinacci, Weinberger, Kelley, Lovell, Diemer, and Hernquist]{nelsonIllustrisTNGSimulationsPublic2019}
Dylan Nelson, Volker Springel, Annalisa Pillepich, Vicente {Rodriguez-Gomez}, Paul Torrey, Shy Genel, Mark Vogelsberger, Ruediger Pakmor, Federico Marinacci, Rainer Weinberger, Luke Kelley, Mark Lovell, Benedikt Diemer, and Lars Hernquist.
\newblock The {{IllustrisTNG}} simulations: Public data release.
\newblock \emph{Computational Astrophysics and Cosmology}, 6\penalty0 (1):\penalty0 2, May 2019.
\newblock ISSN 2197-7909.
\newblock \doi{10.1186/s40668-019-0028-x}.

\bibitem[Ni et~al.(2022)Ni, Di~Matteo, Bird, Croft, Feng, Chen, Tremmel, DeGraf, and Li]{niASTRIDSimulationEvolution2022}
Yueying Ni, Tiziana Di~Matteo, Simeon Bird, Rupert Croft, Yu~Feng, Nianyi Chen, Michael Tremmel, Colin DeGraf, and Yin Li.
\newblock The {{ASTRID}} simulation: The evolution of supermassive black holes.
\newblock \emph{Monthly Notices of the Royal Astronomical Society}, 513\penalty0 (1):\penalty0 670--692, June 2022.
\newblock ISSN 0035-8711.
\newblock \doi{10.1093/mnras/stac351}.

\bibitem[Ni et~al.(2023)Ni, Genel, {Angl{\'e}s-Alc{\'a}zar}, {Villaescusa-Navarro}, Jo, Bird, Di~Matteo, Croft, Chen, {de Santi}, Gebhardt, Shao, Pandey, Hernquist, and Dave]{niCAMELSProjectExpanding2023}
Yueying Ni, Shy Genel, Daniel {Angl{\'e}s-Alc{\'a}zar}, Francisco {Villaescusa-Navarro}, Yongseok Jo, Simeon Bird, Tiziana Di~Matteo, Rupert Croft, Nianyi Chen, Natal{\'i} S.~M. {de Santi}, Matthew Gebhardt, Helen Shao, Shivam Pandey, Lars Hernquist, and Romeel Dave.
\newblock The {{CAMELS Project}}: {{Expanding}} the {{Galaxy Formation Model Space}} with {{New ASTRID}} and 28-parameter {{TNG}} and {{SIMBA Suites}}.
\newblock \emph{The Astrophysical Journal}, 959\penalty0 (2):\penalty0 136, December 2023.
\newblock ISSN 0004-637X.
\newblock \doi{10.3847/1538-4357/ad022a}.

\bibitem[Perez et~al.(2018)Perez, Strub, De~Vries, Dumoulin, and Courville]{perezFiLMVisualReasoning2018}
Ethan Perez, Florian Strub, Harm De~Vries, Vincent Dumoulin, and Aaron Courville.
\newblock {{FiLM}}: {{Visual Reasoning}} with a {{General Conditioning Layer}}.
\newblock \emph{Proceedings of the AAAI Conference on Artificial Intelligence}, 32\penalty0 (1), April 2018.
\newblock ISSN 2374-3468, 2159-5399.
\newblock \doi{10.1609/aaai.v32i1.11671}.

\bibitem[Quinn et~al.(1997)Quinn, Katz, Stadel, and Lake]{quinnTimeSteppingNbody1997}
Thomas Quinn, Neal Katz, Joachim Stadel, and George Lake.
\newblock Time stepping {{N-body}} simulations, October 1997.

\bibitem[Schaller et~al.(2024)Schaller, Borrow, Draper, Ivkovic, McAlpine, Vandenbroucke, Bah{\'e}, Chaikin, Chalk, Chan, Correa, {van~Daalen}, Elbers, Gonnet, Hausammann, Helly, Hu{\v s}ko, Kegerreis, Nobels, Ploeckinger, Revaz, Roper, {Ruiz-Bonilla}, Sandnes, Uyttenhove, Willis, and Xiang]{schallerSwiftModernHighly2024}
Matthieu Schaller, Josh Borrow, Peter~W Draper, Mladen Ivkovic, Stuart McAlpine, Bert Vandenbroucke, Yannick Bah{\'e}, Evgenii Chaikin, Aidan B~G Chalk, Tsang~Keung Chan, Camila Correa, Marcel {van~Daalen}, Willem Elbers, Pedro Gonnet, Lo{\"i}c Hausammann, John Helly, Filip Hu{\v s}ko, Jacob~A Kegerreis, Folkert S~J Nobels, Sylvia Ploeckinger, Yves Revaz, William~J Roper, Sergio {Ruiz-Bonilla}, Thomas~D Sandnes, Yolan Uyttenhove, James~S Willis, and Zhen Xiang.
\newblock Swift: A modern highly parallel gravity and smoothed particle hydrodynamics solver for astrophysical and cosmological applications.
\newblock \emph{Monthly Notices of the Royal Astronomical Society}, 530\penalty0 (2):\penalty0 2378--2419, May 2024.
\newblock ISSN 0035-8711.
\newblock \doi{10.1093/mnras/stae922}.

\bibitem[Stumm and Walther(2010)]{stumm2010new}
Philipp Stumm and Andrea Walther.
\newblock New algorithms for optimal online checkpointing.
\newblock \emph{SIAM Journal on Scientific Computing}, 32\penalty0 (2):\penalty0 836--854, 2010.
\newblock \doi{10.1137/080742439}.

\bibitem[Um et~al.(2021)Um, Brand, Yun, Fei, Holl, and Thuerey]{umSolverintheLoopLearningDifferentiable2021}
Kiwon Um, Robert Brand, Yun, Fei, Philipp Holl, and Nils Thuerey.
\newblock Solver-in-the-{{Loop}}: {{Learning}} from {{Differentiable Physics}} to {{Interact}} with {{Iterative PDE-Solvers}}, January 2021.

\bibitem[{Villaescusa-Navarro} et~al.(2021){Villaescusa-Navarro}, {Angl{\'e}s-Alc{\'a}zar}, Genel, Spergel, Somerville, Dave, Pillepich, Hernquist, Nelson, Torrey, Narayanan, Li, Philcox, La~Torre, Delgado, Ho, Hassan, Burkhart, Wadekar, Battaglia, Contardo, and Bryan]{villaescusa-navarroCAMELSProjectCosmology2021}
Francisco {Villaescusa-Navarro}, Daniel {Angl{\'e}s-Alc{\'a}zar}, Shy Genel, David~N. Spergel, Rachel~S. Somerville, Romeel Dave, Annalisa Pillepich, Lars Hernquist, Dylan Nelson, Paul Torrey, Desika Narayanan, Yin Li, Oliver Philcox, Valentina La~Torre, Ana~Maria Delgado, Shirley Ho, Sultan Hassan, Blakesley Burkhart, Digvijay Wadekar, Nicholas Battaglia, Gabriella Contardo, and Greg~L. Bryan.
\newblock The {{CAMELS}} project: {{Cosmology}} and {{Astrophysics}} with {{MachinE Learning Simulations}}.
\newblock \emph{The Astrophysical Journal}, 915\penalty0 (1):\penalty0 71, July 2021.
\newblock ISSN 0004-637X, 1538-4357.
\newblock \doi{10.3847/1538-4357/abf7ba}.

\bibitem[Wang et~al.(2009)Wang, Moin, and Iaccarino]{wang2009minimal}
Qiqi Wang, Parviz Moin, and Gianluca Iaccarino.
\newblock Minimal repetition dynamic checkpointing algorithm for unsteady adjoint calculation.
\newblock \emph{SIAM Journal on Scientific Computing}, 31\penalty0 (4):\penalty0 2549--2567, 2009.
\newblock \doi{10.1137/080727890}.

\end{thebibliography}
\section*{Acknowledgments}
AT thanks Adrian Bayer and Matthew Ho for helpful discussions, and Alexandre Refregier for his backing of the project.
FL thanks Chihway Chang and Yuuki Omori for early conversations, and the Flatiron Institute for its support.
This work was supported by an ETH Zurich Doc.Mobility Fellowship, which funded AT's research visit to the Flatiron Institute’s Center for Computational Astrophysics in New York. 
TT acknowledges funding from the Swiss National Science Foundation under the Ambizione project PZ00P2\_193352.

\appendix
\crefalias{section}{appendix}
\counterwithin{figure}{section}

\section{Reference simulation data}\label{sec:camels}
Because our approach treats the pressure field $P$ in \cref{eq:ode} as a data-driven component without explicit physical assumptions, we require a ground truth reference to train it.
We employ the fully hydrodynamical Cosmology and Astrophysics with MachinE Learning Simulations (\camels) suite~\citep{villaescusa-navarroCAMELSProjectCosmology2021} for this purpose.

We utilize the cosmic variance (CV) subsets of \camels, which comprise 27 simulations with varying random initial conditions at $z = 127$ while maintaining fixed fiducial hydrodynamical and \lcdm cosmological parameters: $\ocdm = 0.251$, $\obary = 0.049$, $\sigeight = 0.8$, $h = 0.6711$, and $\ns = 0.9624$.
Each simulation evolves $256^3$ dark matter particles in a periodic box of comoving volume $(25\,\mathrm{Mpc}/h)^3$.

Since our neural pressure model is particle-based following the dynamics of \cref{eq:ode}, we focus on hydrodynamical codes that maintain approximately constant gas particle counts and masses throughout the simulation.
This requirement excludes mesh-based codes like \illustris~\citep{nelsonIllustrisTNGSimulationsPublic2019} and restricts our analysis to the mesh-free \simba~\citep{daveSimbaCosmologicalSimulations2019} and smoothed-particle hydrodynamics \astrid~\citep{birdASTRIDSimulationGalaxy2022,niASTRIDSimulationEvolution2022} subsets of \camels. 
For this work, we focus on the \simba simulations, as the strong feedback processes at the fiducial parameters in its CV set provide a more challenging test for our method.
We also tested the neural pressure model on \astrid, but the weak feedback in its CV set produces minimal differences between the gravity-only solution and one including gas physics at our PM resolution.

The \simba simulations are stored as 34 snapshots ($s \in [0,33]$ in \cref{eq:loss}) spanning cosmological redshifts $z = 6$ to the present day ($z=0)$, with scale factor $a(t) = 1/(1 + z)$.
For the results presented in this work, we train our models using only a subset of four ($a_s \in \{0.30, 0.44, 0.65, 1.0\}$) of these snapshots from a single simulation.

\section{Neural pressure implementation details}\label{sec:network_implementation}
We implement neural networks in \flax\footnote{\url{https://flax.readthedocs.io/en/latest/}} and train them using \optax\footnote{\url{https://optax.readthedocs.io/en/latest/}}.
Our default fully convolutional architecture incorporates feature-wise linear modulation (FiLM) conditioning~\citep{perezFiLMVisualReasoning2018} on the scale factor $a$, residual connections~\citep{heDeepResidualLearning2016}, layer normalization~\citep{baLayerNormalization2016}, and circular padding to respect the periodic boundary conditions of the simulation box.
Throughout this work, we use a network comprising 6 hidden layers with 16 channels each and (3,3,3) kernels with unit stride, totaling $47 \, 265$ trainable parameters.
We optimize using Adam~\citep{kingmaAdamMethodStochastic2017} with a constant learning rate of $10^{-4}$ for $1 \, 000$ steps and gradient clipping at global norm 1.
Training stability is generally sensitive to these choices, which we identified via random hyperparameter search.
In the loss $\mathcal{L}$ from \cref{eq:loss}, we set $\lambda = 0.01$, $\mu = 0.1$, $\delta = M/8$, and $\delta' = 4 \delta$, where $M$ is the mesh resolution (number of cells per dimension).

We integrate \cref{eq:ode} over 67 fixed steps in $a$ that exactly include the reference scale factors $a_s$.
To enforce non-negativity and reduce dynamic range, the network predicts $\log{U_\varphi}$ in \cref{eq:pressure}.
During training, we apply random 90° rotations and axis-aligned flips as data augmentation to encourage approximate equivariance to these symmetries.

\section{EGD implementation details}\label{sec:egd_implementation}
In the EGD method, the effective temperature field $T(\delta) = T_0(1 + \delta)^{(\gamma -1)}$ is smoothed by a Gaussian kernel $\hat{\mathbf{O}}(k) = \exp(-(k r_J)^2/2)$ in Fourier space before enthalpy gradients are computed.
We determine the free parameters $T_0 = 0.012$, $\gamma = 1.034$, and $r_J = 8.127$ by optimizing only the power spectrum term from \cref{eq:loss}.
This restriction is necessary because the randomly sampled dark matter particles serving as gas particle proxies lack direct counterparts with particles in the reference hydrodynamical simulation, preventing evaluation of the position and velocity loss terms.

\section{Additional figures}\label{sec:additional_figures}
\begin{figure}[htb]
    \centering
    \includegraphics[width=\linewidth]{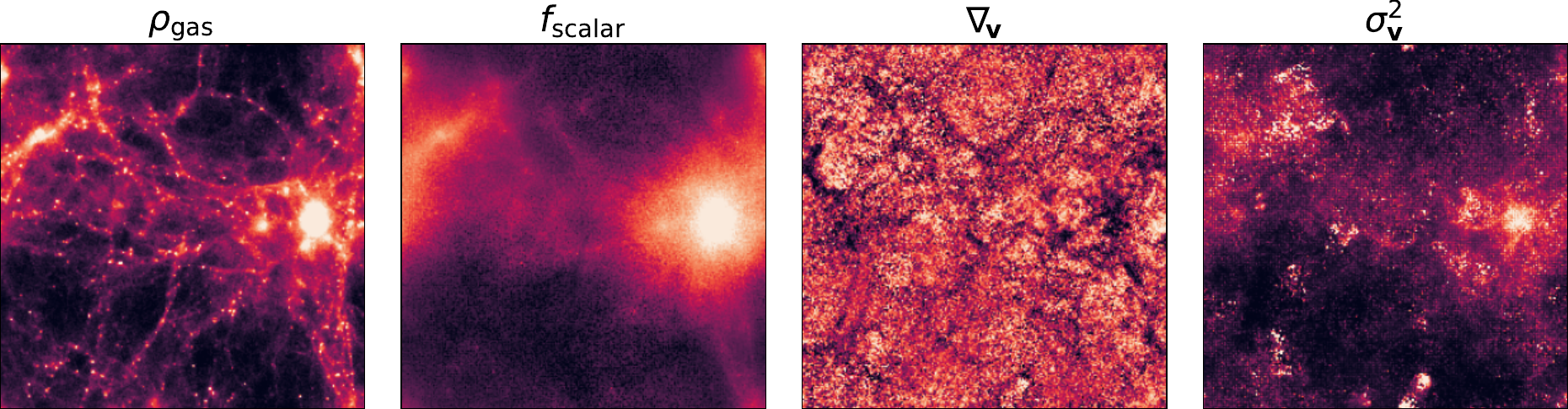}
    \caption{
        Projections of the four local features $\mathbf{h}(\mathbf{x})$ used to predict the effective pressure $P_\varphi$:
        gas density, scalar force (see \cref{eq:fscalar}), velocity divergence, and velocity dispersion (left to right).
        For visualization, each field $h$ is separately scaled by $\mathrm{arcsinh}(h/\sigma_h)$.
    }
    \label{fig:hpm_inputs}
\end{figure}

\begin{figure}[htb]
    \centering
    \includegraphics[width=\linewidth]{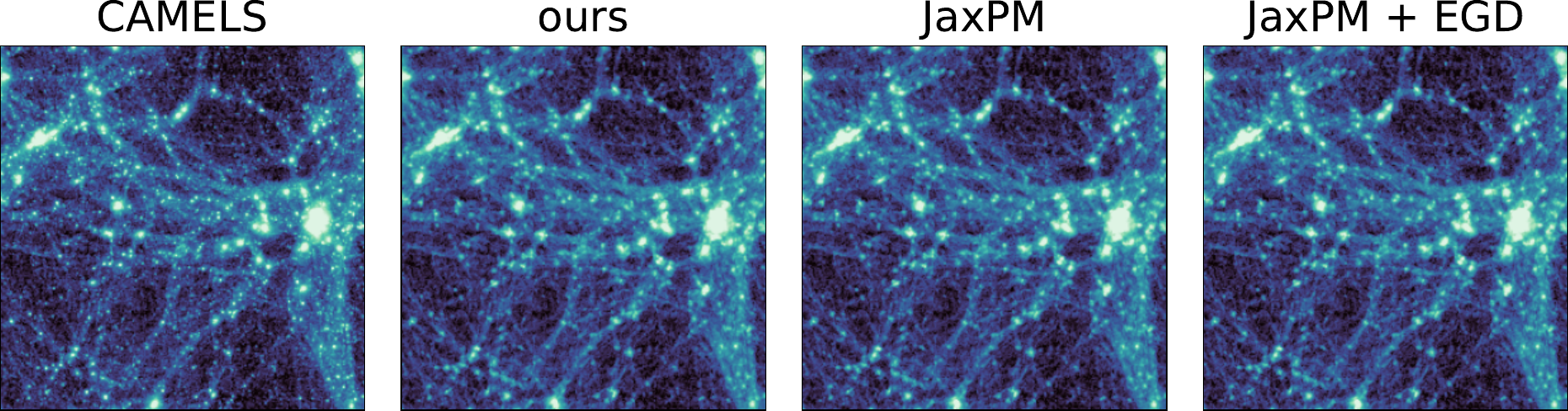}
    \caption{
        Like \cref{fig:final_fields}, but for the dark matter density $\rho_\mathrm{dm}$.
        At this resolution, differences between methods are minimal, demonstrating that our neural pressure model does not adversely impact dark matter evolution.
        The large-scale structure resembles that in \cref{fig:final_fields} because gas traces the underlying dark matter distribution up to scales where gas feedback effects become significant.
    }
    \label{fig:appendix_dm}
\end{figure}

\begin{figure}[htb]
    \centering
    \includegraphics[width=\linewidth]{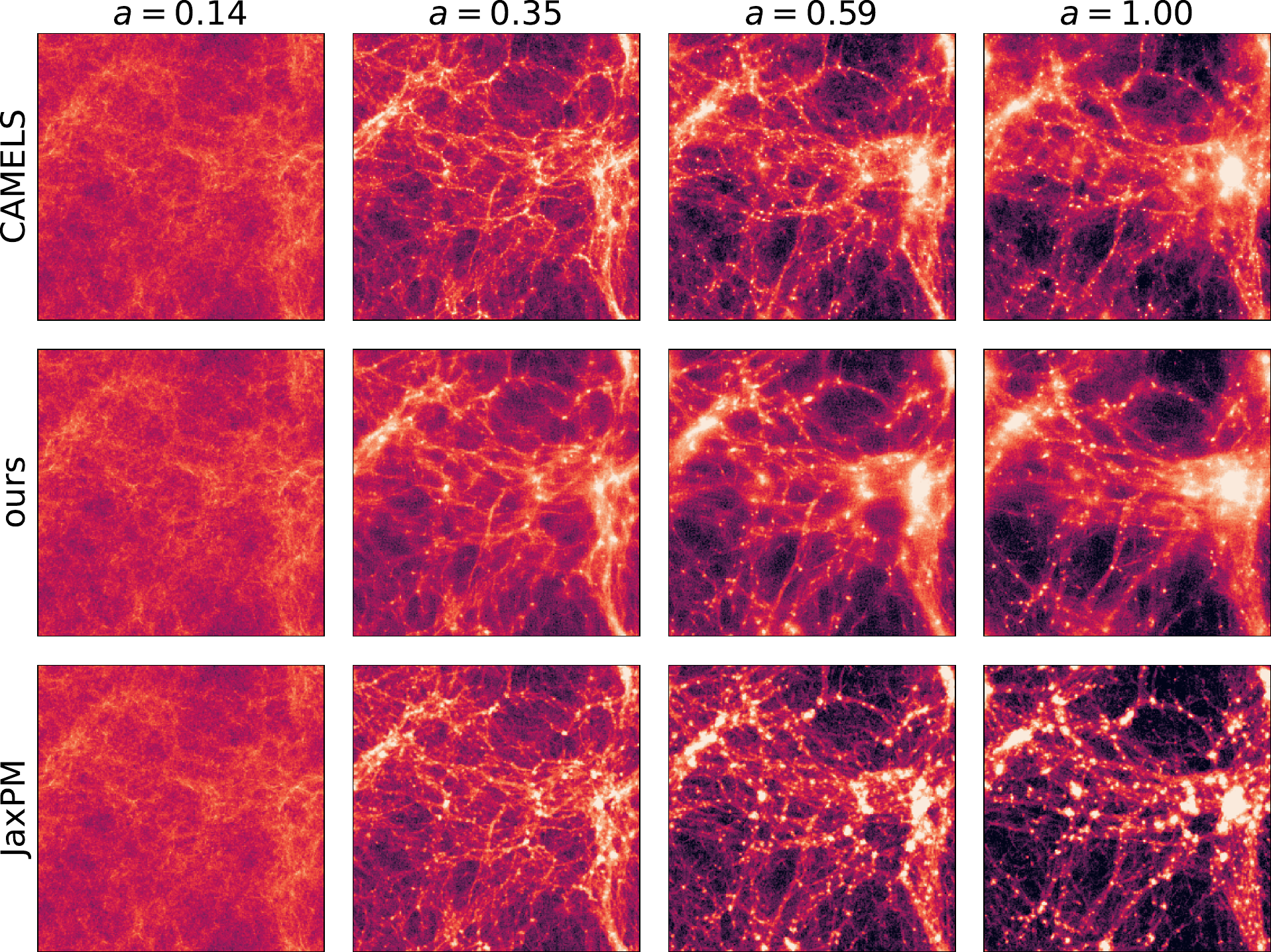}
    \caption{
        Evolution of gas density $\rho_\mathrm{gas}$ from our initial conditions ($a=0.14$) to the present day ($a = 1$).
        For this example, the penalized scale factors in the loss function \cref{eq:loss} are $a_s \in \{0.30, 0.44, 0.65, 1.0\}$.
        As a post-processing method, EGD does not model a self-consistent evolution and is therefore not shown here.
    }
    \label{fig:appendix_evolution}
\end{figure}

\end{document}